\documentclass{article}

\usepackage{arxiv}

\usepackage[utf8]{inputenc} 
\usepackage[T1]{fontenc}    
\usepackage{hyperref}       
\usepackage{url}            
\usepackage{booktabs}       
\usepackage{amsfonts}       
\usepackage{nicefrac}       
\usepackage{microtype}      
\usepackage{lipsum}		
\usepackage{graphicx}
\usepackage{natbib}
\usepackage{doi}

\title{Analysis of the relation between smartphone usage changes during the COVID-19 pandemic and usage preferences on apps}

\author{ \href{https://orcid.org/0000-0002-5501-7374}{\includegraphics[scale=0.06]{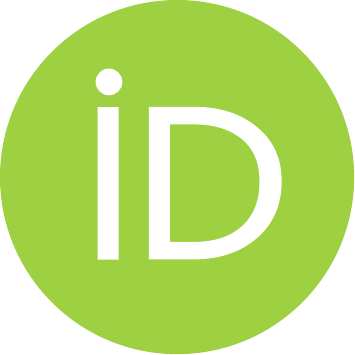}\hspace{1mm}Yuxuan Yang}\\
	 Graduate School of Science and Technology\\
	University of Tsukuba\\
	1 Chome-1-1 Tennodai, Tsukuba \\
	\texttt{s2030112@s.tsukuba.ac.jp} \\
	\And
	\href{https://orcid.org/0000-0002-3671-9434}{\includegraphics[scale=0.06]{orcid.pdf}\hspace{1mm}Maiko Shigeno} \\
	Faculty of Engineering, Information and Systems\\
	University of Tsukuba\\
	1 Chome-1-1 Tennodai, Tsukuba \\
	\texttt{maiko@sk.tsukuba.ac.jp} \\
}

\renewcommand{\shorttitle}{\textit{arXiv} }

\begin{document}
\maketitle

\begin{abstract}

Since the World Health Organization announced the COVID-19 pandemic in March 2020, curbing the spread of the virus has become an international priority. It has greatly affected people's lifestyles. 
In this article, we observe and analyze the impact of the pandemic on people's lives using changes in smartphone application usage.
First, through observing the daily usage change trends of all users during the pandemic, we can understand and analyze the effects of restrictive measures and policies during the pandemic on people's lives. In addition, it is also helpful for the government and health departments to take more appropriate restrictive measures in the case of future pandemics. 
Second, we defined the usage change features and found 9 different usage change patterns during the pandemic according to clusters of users and show the diversity of daily usage changes. It helps to understand and analyze the different impacts of the pandemic and restrictive measures on different types of people in more detail.
Finally, according to prediction models, we discover the main related factors of each usage change type from user preferences and demographic information.
It helps to predict changes in smartphone activity during future pandemics or when other restrictive measures are implemented, which may become a new indicator to judge and manage the risks of measures or events.
\end{abstract}


\section{Introduction}
Since the World Health Organization declared the COVID-19 outbreak a pandemic on March 11, 2020, curbing the spread of the virus has become an international priority \cite{gao2020mental}. To reduce the spread of the COVID-19 pandemic, many countries or regions around the world have successively adopted a series of measures and policies to restrict people's movement. It has had an unprecedented impact on the lives and psychological conditions of countless people. 


In Japan, since the epidemic began to spread, the Japanese government implemented the first Declaration of a State of Emergency for the COVID-19 pandemic in major cities on April 7, 2020, requiring residents to cooperate to prevent infection, including avoidance of nonnecessary going out, suspension of school, and telecommuting for work. The declaration was extended to the whole country on April 16, 2020 \cite{COVID19I85:online}.
The Japanese government has requested people to avoid situations that increase the risk of infection but has not restricted activities by strong strategies such as lockdowns in some towns. Thus, monitoring the activity of people from various data sources became important. Monitoring information also helps people understand the pandemic situation and mind the risk of infection. The government updates basic information daily, such as the number of newly confirmed cases and the number of people requiring inpatient treatment\cite{kannsennninnzu}. 
Data on the number of subway passengers are shown to understand the effect of stay-at-home measures \cite{Toeisubw4:online}.
 Based on a route selection log of users on a route-searching smartphone app, simulating people's movement by trains is useful to avoid congestion \cite{okano2021}.
Using the data collected by the chatbot-based health care systems, epidemiological trends could be monitored\cite{yoneoka2020large}.
Data on people's behaviors play an important role in observing and suppressing pandemics.

Except for monitoring the overall situation during a pandemic, Sallis et al.\cite{sallis2000behavioral} claimed that understanding the changes in individuals' behaviors after being affected by physical and mental health symptoms or other interventions is also helpful to limit the spread of pandemics. 
Philip et al. \cite{philip2020psychology} suggested that a deep understanding of human behavior during pandemics can be helpful for creating better restrictive measures. 
Many studies have investigated many aspects of people's behavioral changes during the pandemic based on surveys.
Gao et al. \cite{gao2020mental} indicated that the high prevalence of mental health problems during the COVID-19 outbreak is positively associated with frequent social media exposure through analyzing questionnaire results.
Di et al. \cite{di2020eating} and Kumari et al. \cite{kumari2020short} showed the changes in people's eating habits and living habits during the lockdown period in Italy and India, respectively, through questionnaires.
Alzueta et al. \cite{alzueta2021covid} explained the impact of the COVID-19 pandemic and subsequent social restraints and quarantines on the mental health of the global adult population by analyzing data collected by an online survey.
Agarwal et al. \cite{agarwal2021global} implemented an online questionnaire platform to collect survey-based data and indicated that there was a moderate to severe effect on peoples' social, financial, and mental health conditions during the COVID-19 disease outbreak.
Although questionnaire surveys might provide enough information we need, it can be difficult to survey in time with each phase of a rapidly changing pandemic such as COVID-19. 
Therefore, we investigate the possibility of understanding people's behavioral changes using data indicating one aspect of people's behavior.

Due to the popularity of smartphones and advances in related technologies, people can mine information from smartphone use. Lv et al. \cite{lv2013mining} presented an approach for mining the users' regular, daily activities based on trajectories of smartphones in the long term and found a similar relationship between users according to their regular use patterns. 
An increasing number of functions can be handled by smartphones, and people's lifestyles and habits have been significantly reflected in the diversification and individualization of using smartphones. Jones et al. \cite{jones2015revisitation}, Banovic et al. \cite{banovic2014proactivetasks}, and Zhao et al. \cite{zhao2016discovering}, proved that diverse user types can be found according to the user's actions performed on the emails displayed on the lock screen, the app revisitation patterns, the preferences for time and categories, and other application usage behaviors.
In addition, many studies have proposed characterizing and predicting users' behavior and personality based on their smartphone usage, which can be used for personalized service identification, recommendation systems, etc.
Mafrur et al. \cite{mafrur2015modeling} proposed an approach to modeling human behavior based on user smartphone data logs for identification.
Sarker et al. \cite{sarker2019recencyminer} mined recent, personalized behavioral patterns based on rules for an individual's contextual phone log data to build an effective personalized usage behavior prediction model.
De Montjoye et al. \cite{de2013predicting} claimed that they can predict users Big Five personality types reliably with the proposed method according to smartphone logs.


In this study, we discuss a possible method for explaining people's behavioral changes based on daily smartphone use during the COVID-19 pandemic.
This study has three main contributions. First, we analyze the impact of self-restraint on people's behavior according to the overall situation using the changes in smartphone usage during the pandemic.
Second, we discovered 9 diverse change patterns in smartphone usage during the pandemic and analyzed how the pandemic affects users with different change patterns.
Third, the main factors related to various usage change patterns were examined from demographic information and smartphone usage preferences in the early stage of the pandemic.

\section{Data set}
The data were provided by Fuller, Inc., which is an application analysis support company that conducts joint research as an industry-university cooperation project.
These data are collected from Android OS smartphones using the premise of obtaining the user's permission based on sufficient instructions.
In our study, access logs and demographic information were used as data. In addition, since our study focused on the situation in Japan, the data whose locale were recorded as Japan were used.

The access log is data recorded at the moment of accessing each device and which application was accessed. It is represented as logs including the device ID, access date and time, and the app ID. An example of access log data is shown in Table \ref{accesslog}. The device ID is a random value given by Fuller, Inc., which is used to uniformly identify the device. The app ID is the information that can identify the name and category of the application on Google Play.

\begin{table}[!ht]
\centering
\caption{{\bf Example of access log data}}
\begin{tabular}{|l|l|l|}
\hline
device ID & datetime            & app ID                \\ \hline
device1 ID & 2020-01-31T15:37:03 & com.twitter.android   \\ \hline
device2 ID & 2020-01-31T19:20:14 & jp.naver.line.android \\ \hline
...      & ...                 & ...                   \\ \hline
\end{tabular}
\label{accesslog}
\end{table}

The demographic information data included the users' gender and birth year. An example is shown in Table \ref{demo}.  The device ID in demographic information can be combined with the device ID in the access log.
\begin{table}[!ht]
\centering
\caption{{\bf Example of demographic information data}}
\begin{tabular}{|l|l|l|}
\hline
devide ID  & gender         & birth year             \\ \hline
devide1 ID & female & 1994  \\ \hline
devide2 ID & male & 1988\\ \hline
...      & ...                 & ...                   \\ \hline
\end{tabular}
\label{demo}
\end{table}

\subsection*{Definition of target periods}
\begin{figure}[!ht]
\includegraphics[scale=0.32]{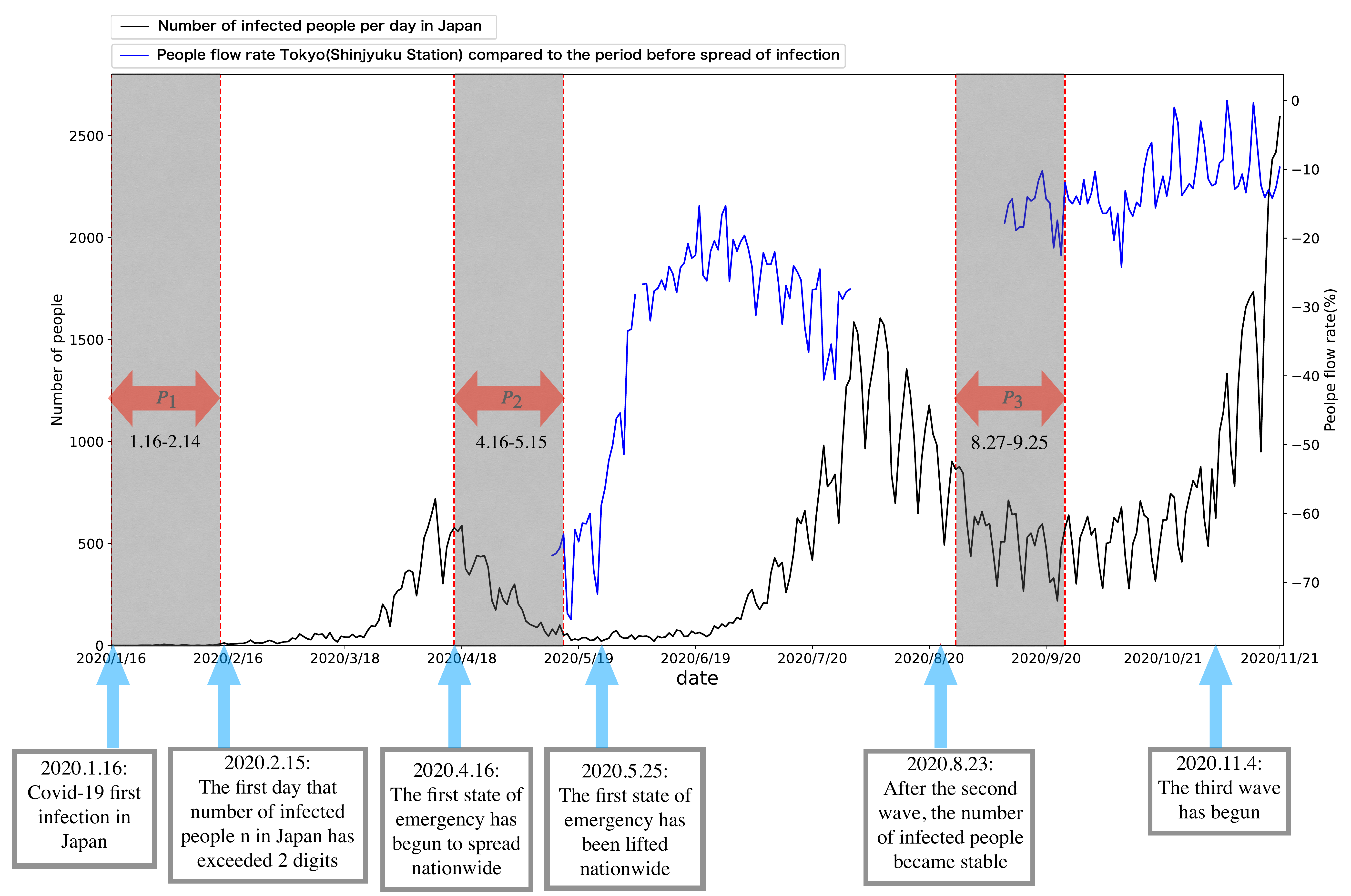}
\centering
\caption{{\bf The daily number of infections nationwide and the flow rate of Tokyo}}
\label{Fig1}
\end{figure}

To observe the changes in smartphone usage during the pandemic, we selected three periods to be analyzed.
First, the number of people confirmed as COVID-19 positive per day in Japan increased to double digits on February 15, 2020, for the first time. Thus, the previous month, i.e., January 16, 2020--February 14, 2020, was selected as the pre-pandemic period.
Then, according to the time and date when Japan's first Declaration of a State of Emergency was implemented nationwide on April 16, 2020, the month after that, i.e., April 16, 2020--May 15, 2020, was selected as the self-restraint period.
Finally, the period from the end of August to mid-October was the period when the number of people infected after the second wave of the epidemic in Japan was under control. It was a period that was still part the epidemic but without the restrictions of the Declaration of a State of Emergency.
To align the number of weeks and days with the previous two periods, we selected the period from August 27, 2020, to September 26, 2020. We define these three periods as $P_1, P_2$, and $P_3$, respectively. Fig \ref{Fig1} shows the daily number of people infected in Japan (the data from \cite{kannsennninnzu})and the people flow rate in Tokyo (Shinjuku Station) during the pandemic period compared to the period before infection spread (the data from \cite{COVID19I85:online}).

Since there are different tendencies of daily smartphone usage between weekdays and weekends \cite{deng2019measuring}, we remove weekends (Saturday and Sunday) from each period. Each period is consisting of 22 days. We use the notations $P_1, P_2$ and $P_3$ to represent the set of days in each period. Note that the numbers of national holidays in each period are different as shown in Table \ref{target period}.

\begin{table}[!ht]
\centering
\caption{{\bf Selection of target period}}
\begin{tabular}{ccccc}
\hline
period& range  &state& national holidays \\\hline
$P_1$&  2020.1.16(Thu) --2020.2.14(Thu) &before pandemic& 2.11(Tue) 2.24(Mon) \\
$P_2$& 2020.4.16(Thu) -- 2020.5.15(Thu) &self-restraint&4.28(Wed) 5.4(Mon) 5.5(Tue) 5.6(Wed)  \\
$P_3$& 2020.8.27(Thu) -- 2020.9.25(Thu) &no state of self-restraint&9.21(Mon) 9.22(Tue) \\
\hline
\end{tabular}
\label{target period}
\end{table}

\subsection*{Definition of analysis target apps}

\begin{figure}[!ht]
\includegraphics[scale=1]{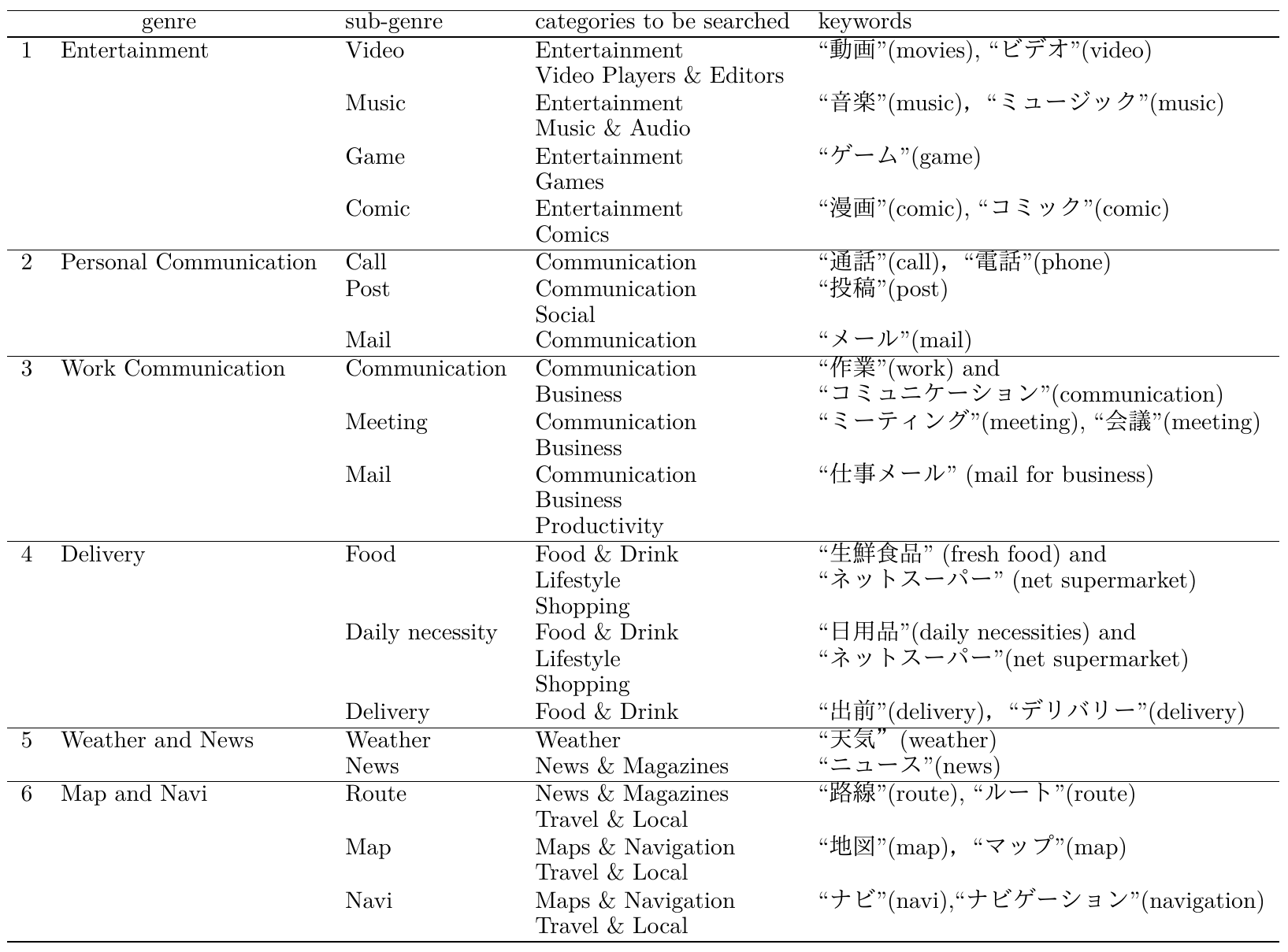}
\centering
\caption{{\bf Genre and its keywords for selecting apps}}
\label{Table4}
\end{figure}

There are many apps for smartphones that differ in the number of app users. 
Thus, we need to select apps targeted for analysis.
To analyze the changes in smartphone usage before and after the epidemic, according to the questionnaire survey \cite{54:online}, we selected 6 app genres with a change ratio of usage time over 20\% during the pandemic and subdivided these 6 genres into 18 subgenres, which are shown in Fig \ref{Table4}.
Then, we selected apps that contained specified keywords in the app description and belonged to the designated app categories in Google Play in March 2021. Fig \ref{Table4} shows the 18 subgenres and app categories and keywords we used for each genre.
The keywords are given by Japanese represented in " ". 
In each subgenre, we sort the selected apps in descending order according to the number of reviews in Google Play. The top 50 apps for each subgenre were selected (select all if less than 50). The number of apps selected by keywords in every subgenre is shown in the third column of Table \ref{genre and sub-genre}.

\begin{table}[!ht]
\centering
\caption{{\bf The count of apps in each genre and sub-genre}}
\begin{tabular}{lllccc}\hline
\multicolumn{2}{c}{genre} & sub-genre &\begin{tabular}{c}\ apps \\selected by \\keywords \end{tabular}&\begin{tabular}{c} \ candidate \\apps  \end{tabular}& \begin{tabular}{c}\ apps to \\be analysed \end{tabular}\\ \hline
1&Entertainment &Video &49&5\\
& & Music&50 &11\\
& & Game &50&13\\
& & Comic &41 &12& 40\\\hline
2&Personal Communication & Call &47 &8\\
& & Post &43&4\\
& & Mail &44&7&15\\ \hline
3&Work Communication &Communication &12& 1 \\
& & Meeting &33&5\\
& & Work mail &49&12&16\\ \hline
4&Delivery &Food &1& 1\\
& & Daily necessity &3&1\\
& & Delivery &22&5&6 \\\hline
5&Weather and News & Weather & 47&3\\
& & News & 49&10& 13\\ \hline
6&Map and Navi & Route &47&5\\
& &Map &47&6 \\
& & Navi & 44&4& 6\\ \hline
\end{tabular}
\label{genre and sub-genre}
\end{table}

Then, we collected the access log data for the selected apps in every subgenre during the target periods, and we sorted the apps for each subgenre in descending order in the count of the users that have used the apps during target periods.
We accumulated the percentage of each app's user count in the sum of the user count of apps in each subgenre in this order and collected applications until the cumulative percentage exceeds 90\% as candidate applications for each subgenre.
The set of candidate apps of subgenre $sg (=1, \dots, 18)$ is denoted by $A_{sg}$. 
The set of the target apps of genre $g$ can be represented as $A_g=\bigcup_{sg \in g } A_{sg}$. 
The number of target apps in each subgenre and each genre is shown in Table \ref{genre and sub-genre}.
We have $\sum_{sg}|A_{sg}|=113 $ and $\sum_{g}|A_{g}|=96$, because some apps belong to several subgenres in the same genre. There are 5 apps belonging to the genres 'Personal Communication' and 'Work Communication' simultaneously, and the total number of target apps is $\bigcup_{g}|A_{g}|=91$.

\subsection*{Definition of analysis target users}
We extracted users whose unique age and gender could be identified in the demographic information and who had activated at least one target app during the target periods.
In addition, we restricted users from 10 to 69 years old.
As a result, 27,004 users were selected for our analysis.
We denote the set of these selected users by $N$.
Table \ref{age and gender} shows the percentage distribution of target users in different age groups and genders. 
The average age of all target users is 36.9. 

\begin{table}[!ht]
\centering
\caption{{\bf Distribution of target users (\%)} \label{age and gender}}
\begin{tabular}{crrrrrrr}\hline
Age group&10s&20s&30s&40s&50s&60s& total\\ \hline
Male&9.10&9.45&6.03&8.46&8.22&4.73&45.99\\
Female&8.90&8.85&11.46&14.35&7.55&3.89&54.00\\
total & 18.00&18.30&17.49&22.81&15.77&7.62&100.0
\\ \hline 
\end{tabular}
\label{age and gender}
\end{table}

\section{Analysis of smartphone usage in daily life}
\subsection*{Feature amount on daily usage}

To compare the smartphone usage trends in daily life during the pandemic, we quantified the daily usage during the target periods. 
We defined daily smartphone usage based on the access log data as follows.
We divided the time of day in target period $P_i$ into 24 time spans. 
The time span $t$ refers to the period from $t:00:00$ to $t:59:59 (t = 0, \dots, 23)$ in the time zone Asia/Tokyo (UTC+9).
By aggregating access log data in 24-hour time spans, we expect to find changes in life rhythm. 
We created a feature vector by counting the number of logs accessing target apps in every time span during each period. 
However, there were large differences in the number of recorded access logs in one use for each app. 
Some apps had many access logs because users usually visited many pages with one use. 
Other apps had a few logs by one use because it was enough for them to visit only a few pages. 
It was difficult to find a trend in daily life due to this variance.
Therefore, we adopted the number of days the apps were accessed as the amount of daily usage, which was more appropriate to the type of data available.
Let $d_i^{n, t}$ be the number of days user $n$ accessed any target apps in $A$ in daily time span $t$ during period $P_i$. We denote the number of accessed days per time span during period $P_i$ as the time series $F_i^n = (d_i^{n, 0}, \dots, d_i^{n, t},  \dots, d_i^{n, 23})$, which represents the daily smartphone usage by user $n$ during period $P_i$.

\begin{figure}[!ht]
\includegraphics[scale=0.6]{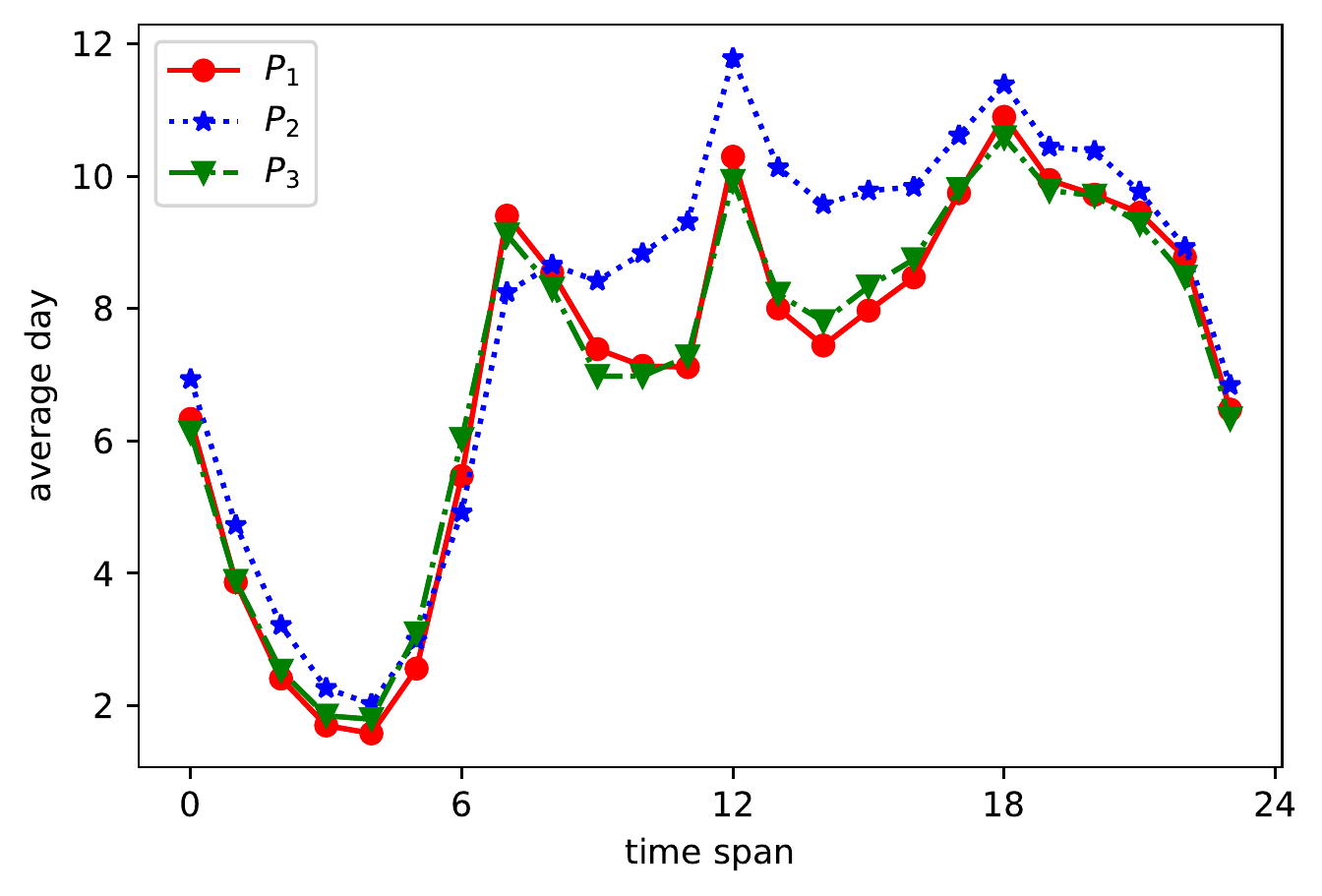}
\centering
\caption{{\bf The average of all users' daily usage features in total genres}}
\label{Fig2}
\end{figure}

To observe daily usage changes during the pandemic, we calculated the average daily usage features by users during the target periods, i.e., $\sum_{n \in N} F_i^{n}/|N|, i=1, 2, 3$ in Fig \ref{Fig2}.
There was less change of the average usages between period $P_1$ and $P_3$, whereas in period $P_2$, usage was pretty different.
The usage increased significantly during $P_2$ compared to $P_1$ and $P_3$. 
In particular, it increased during the time span 9--18, which were the working hours in general.
The usage in $P_2$ decreased during the time span 7--8, which was the commuting hour in general. 
In $P_1$ and $P_3$, we confirmed that there were peaks of daily usage at time spans 8, 12 and 18. However, in $P_2$, the peak at time span 8 vanished.
We can see the impact of self-restraint on daily smartphone usage.

To observe the changes in daily usage trends in different genres, 
we similarly defined the amount of daily usage of each genre by the number of days the apps were accessed.
We denoted the number of days user $n$ accessed any target apps belonging to genre $g$ in the daily time span $t$ during period $P_i$ as $d_i^{n, t, g}$. 
The time series $F_i^{n,g} = (d_i^{n, 0, g}, \dots, d_i^{n, t, g},  \dots, d_i^{n, 23, g})$ was used to represent the amount of daily usage of genre $g$ by user $n$ during period $P_i$.

\begin{figure}[!ht]
\includegraphics[scale=0.6]{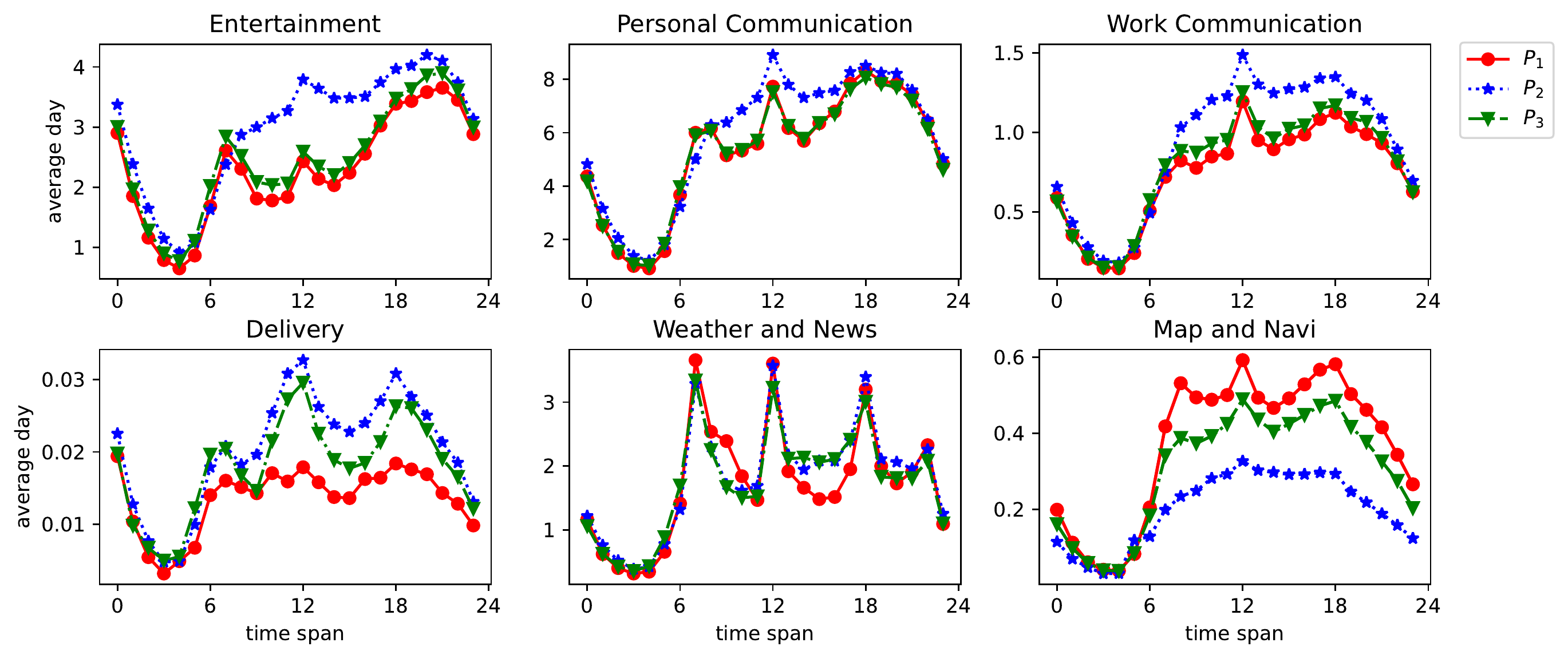}
\centering
\caption{{\bf The average of all users' daily usage features of each genre}}
\label{Fig3}
\end{figure}

We show the average users' daily usage of each genre, i.e., $\sum_{n \in N} F_i^{n,g}/|N|$, $i=1, 2, 3$, $g=1, \dots, 6$, in Fig \ref{Fig3}.
Note that the scales of the vertical axis of these graphs vary.
For the genres of 'Personal Communication', 'Entertainment', and 'Work Communication', compared to $P_1$, the usage in $P_2$ significantly increased, whereas the usage in $P_3$ recovered to the same level as $P_1$. 
This kind of change can be considered to indicate that working remotely was fulfilling, and flexible hours were realized for many users during the self-restraint period.
For the genre 'Map and Navi', compared to $P_1$, the usage in $P_2$ significantly decreased, and the usage in $P_3$ had
a tendency to increase, but it still had not returned to the level in $P_1$. This tendency was interpreted as self-restraint in people's movement.
The usage of the genre 'Delivery' grew in $P_2$, and it declined in $P_3$ but still maintained a high level of usage.
This change could be interpreted as growth in delivery apps during self-restraint.
The usage of the genre 'Weather and News' did not obviously change.


\subsection*{Discovering usage change patterns based on clustering users}

To discover the effects of the pandemic and self-restraint on different users, 
we quantified the changes in daily usage during the pandemic.
We defined the change in the daily usage from period $P_{i-1}$ to the next period $P_i (i =  2, 3)$ by the feature vector 
\[\Delta{F_i^n} = (d_i^{n, 0} -d_{i-1}^{n, 0}, d_i^{n, 1} - d_{i-1}^{n, 1}, \dots, d_i^{n, 23} - d_{i-1}^{n, 23}) \]
This feature vector $\Delta{F_i^n} $ refers to the usage feature change using all target apps.

According to this feature vector $\Delta{F_i^n} $, we defined the dissimilarity among users $n$ and $n'$   by $\| \Delta{F_i^n} -\Delta{F_i^{n'}}  \|_2$, i.e., the Euclidean distance between $\Delta{F_i^n}$ and $\Delta{F_i^{n'}}$, and grouped users based on this dissimilarity using the k-means method.
We found two types of clusters of users based on $\Delta{F_2^n}$ and $\Delta{F_3^n}$, i.e., the changes in daily use from $P_1$ to $P_2$ and $P_2$ to $P_3$. The number of clusters of each type was set to 3. The selection method for the initial value was k-means++, which avoids the effect of initial values. The maximum number of iterations of the k-means method was limited to 10,000. 

\begin{figure}[!ht]
\includegraphics[scale=1]{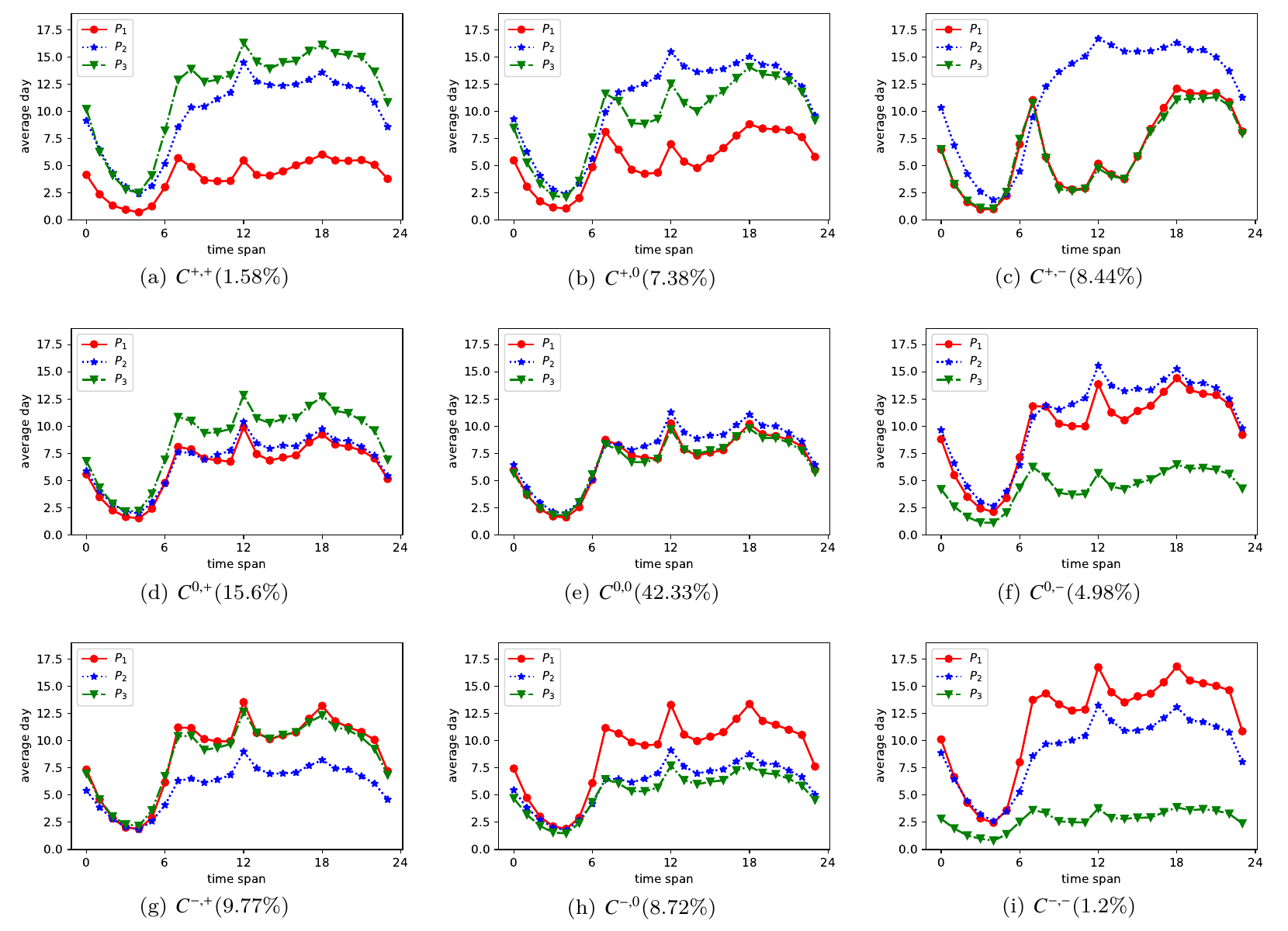}
\centering
\caption{{\bf The average usage in all genres of the 9 clusters with different change features during period $P_1$, $P_2$, and $P_3$}}
\label{Fig4}
\end{figure}

To overview the characteristics of each cluster during the target periods, we grouped 9 clusters obtained from crosses of two types of clusters based on the changes from  $P_1$ to $P_2$ and on the changes $P_2$ to $P_3$, respectively. 
We show the average use during the target periods for users in each cluster $C$, i.e., $\sum_{n \in C} F_i^{n}/|C|, i=1, 2, 3$ in Fig \ref{Fig4}. 
The proportion of users in each cluster to all target users is shown after the name of each cluster in Fig \ref{Fig4}.
We can clearly observe that these 9 clusters have different change patterns and usage features during the 3 target periods. Each row in this figure has different change patterns from $P_1$ to $P_2$, and each column has different change patterns from $P_2$ to $P_3$. 
We can see that the usage from $P_1$ to $P_2$ in the first row increased sharply over almost all time spans. This pattern can be explained by the clusters in which the chances and situations for using smartphones increased significantly during self-restraint period, $P_2$ compared to period $P_1$. 
The use in the second row was interpreted as the clusters that had few changes to use smartphones during $P_1$ and $P_2$, and the use in the third row was significantly reduced during the entire daytime, but it remained unchanged during the late night. The usage during the work time (from time span 5 to 23) from $P_2$ to $P_3$ in the first column increased. This finding can be interpreted as clusters with more use after self-restraint than during self-restraint. The usage from $P_2$ to $P_3$ in the second column decreased slightly or had few changes. The usage from $P_2$ to $P_3$ in the third column dropped sharply throughout the day.



For convenience in the following description, according to the change trend from $P_1$ to $P_2$, we named the clusters in the first, second, and third rows $C^{+, \bullet}, C^{0 ,\bullet}$ and $C^{- ,\bullet}$, respectively. We also named the clusters in the first, second, and third columns $C^{\bullet, +}, C^{\bullet, 0}$ and $C^{\bullet, -}$, respectively, according to the change trend from $P_2$ to $P_3$.
Combining these notations for rows and columns, we distinguished 9 clusters. For instance, $C^{+,+}$ is the cluster represented in the upper left, and $C^{0, -}$ is the cluster represented in the second row and the third column.

\begin{figure}[!ht]
\includegraphics[scale=0.7]{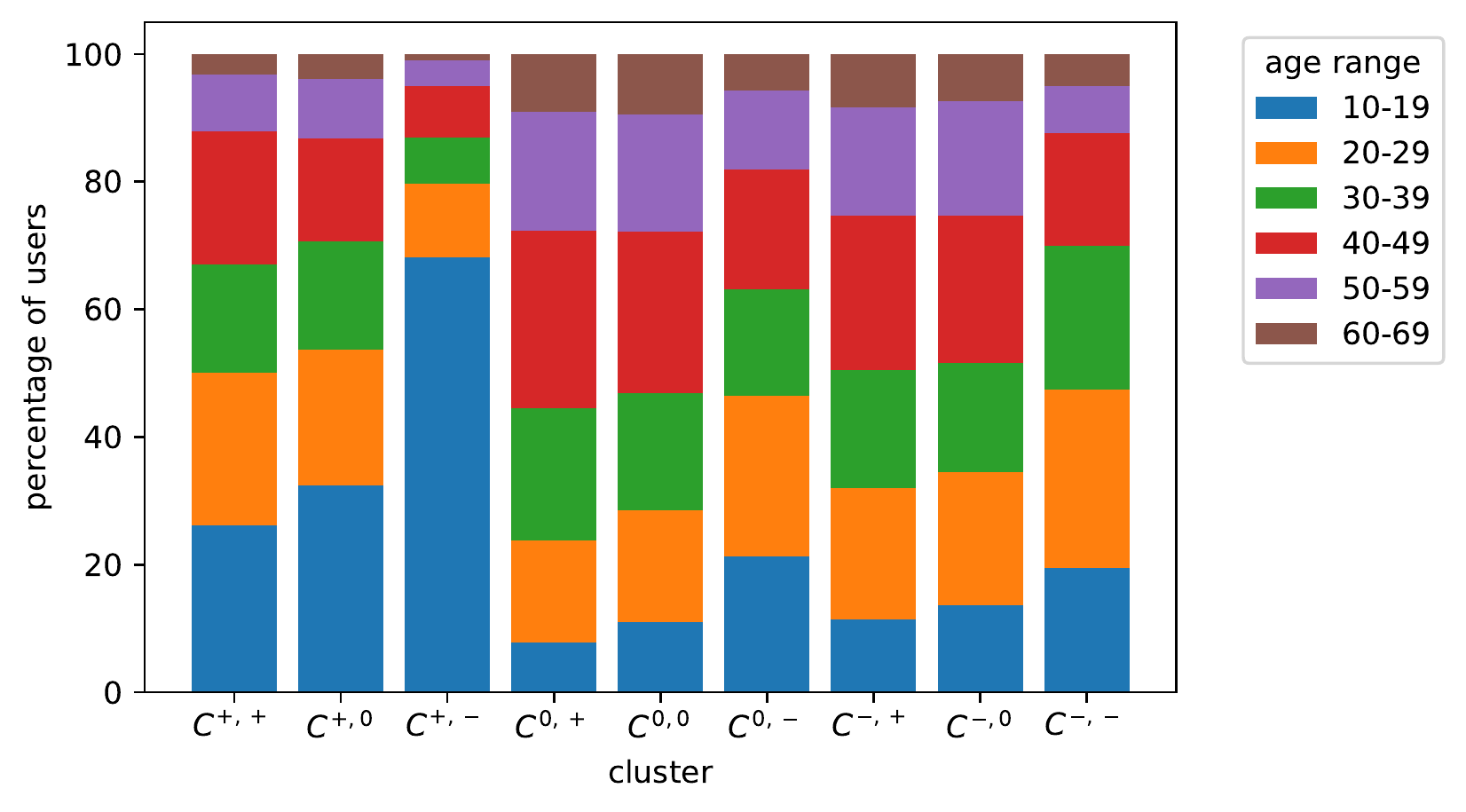}
\centering
\caption{{\bf Age distribution of users in each cluster}}
\label{Fig5}
\end{figure}

\begin{figure}[!ht]
\includegraphics[scale=1]{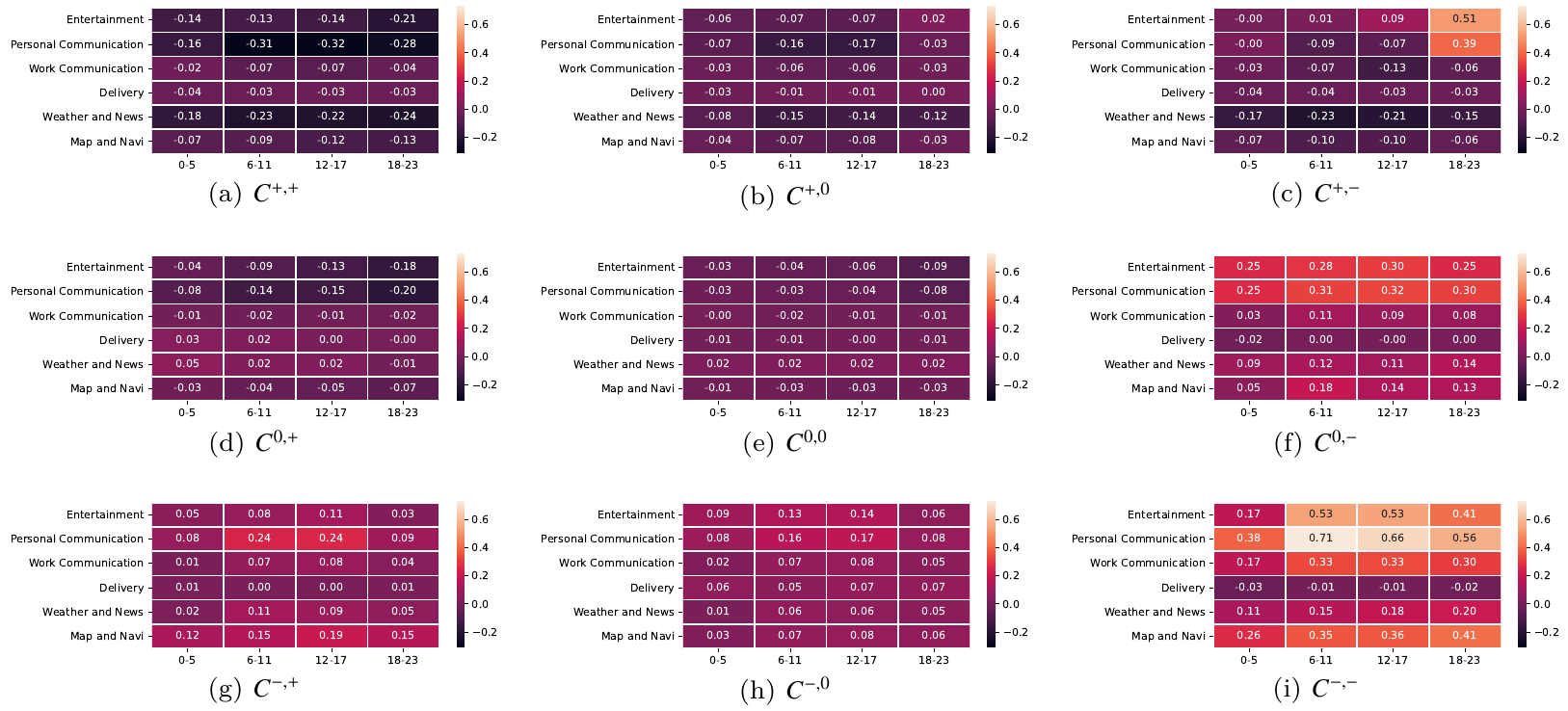}
\centering
\caption{{\bf The average usage in all genres of the 9 clusters with different change features during period $P_1$, $P_2$, and $P_3$}}
\label{Fig6}
\end{figure}

We observed the characteristics of each cluster using Figs \ref{Fig5} and \ref{Fig6} together with Fig \ref{Fig4}. 
Fig \ref{Fig5} displays the histogram of the percentage of each age range of users belonging to each cluster.
Fig \ref{Fig6} shows the average daily usage of users in each cluster for 6 genres at 4 time spans during the pre-pandemic period $P_1$.
Here, for daily usage, we defined $a^{g, n, t}$ as the count of accessing any target apps in $A_g$ by user $n ( \in N)$  in time span $t ( = 0, \dots, 23)$ during period $P_1$. 
The daily usage during period $P_1$ of user $n$ is represented by $(\sum_{t=0}^5 a^{g, n, t}, \sum_{t=6}^{11} a^{g, n, t}, \sum_{t=12}^{17} a^{g, n, t}, \sum_{t=18}^{23} a^{g, n, t})$ for $g=1, \dots, 6$.
These data were standardized. Since we supposed that usage changes were related to daily usage during the pre-pandemic period, we employed this daily usage of apps only during $P_1$.

The largest cluster, $C^{0, 0}$, includes approximately 43\% of the target users. Their usage changes were the smallest among the 9 clusters. The use from $P_1$ to $P_2$ tended to increase slightly during the time span 7--22. 
However, because the magnitude of the change was very small, we believe that these users were people whose use of smartphones was less affected by the pandemic or self-restraint. Because the proportion of this cluster was large, there was no significant difference in the usage distribution by genre or time span from Fig \ref{Fig6}. However, the distribution of the age ranges tended to be slightly biased to the middle age range, as shown in Fig \ref{Fig5}.

The cluster with the second-highest proportion was $C^{0, +}$, whose users accounted for approximately 16\% of the target users. 
The use of smartphones during $P_2$ was not greatly affected by the self-restraint policy, but they entered a more active use pattern during $P_3$. We considered that they had changed, such as increasing online activity due to the epidemic. Similar to $C^{0, 0}$, this cluster also consisted of a slightly larger proportion of the middle age range.

Each of the proportions of  $C^{-,+}$, $C^{-, 0}$, and $C^{+, -}$ was slightly  less than 10\%. The users in $C^{-, +}$ had fewer chances for smartphone use during $P_2$ compared with $P_1$, but the use in $P_3$ was almost the same as in $P_1$. The users in $C^{-, 0}$ also had fewer chances to use smartphones during $P_2$ than in $P_1$ and have entered a more sluggish state of use during $P_3$. Although these changes in usage patterns in $C^{-, +}$ and $C^{-, 0}$ were not very large, the phenomenon that morning peaks of use were small during periods when the use decreased was obvious. This finding can be interpreted as users who had fewer chances to use smartphones during the morning commute when usage was reduced.
The usage pattern in $C^{+, -}$ changed substantially from $P_1$ to $P_2$. The exceptional increase in usage at time spans 7--23 caused the peak usage in the morning and evening to disappear, but the usage peak at approximately noon still existed. It was surprising that the use patterns between $P_1$ and $P_3$ were almost the same. 
We supposed that the self-restraint policy greatly influenced smartphone usage in this cluster. Users might use smartphones freely during the day and late at night because they might not need to commute in the early morning by changing their work or study habits to staying at home. As shown in Fig \ref{Fig6}, the users belonging to this cluster had extremely high entertainment and personal communication usage in the nighttime period 18--23. Moreover, as shown in Fig. 5, this cluster represented the largest proportion of teens.

The users in cluster $C^{+, 0}$ accounted for approximately 7\% of the target users had a different usage pattern between $P_2$ and $P_3$, but it belonged to a cluster with a similar usage pattern between these periods. Like this cluster, some usage patterns in $P_2$ had no peak in the morning, such as commuting hours.
The remaining three clusters had small proportions of users. 
These clusters had distinctive features in average daily usage shown in Fig \ref{Fig6}. 

In addition, there was a trend that the clusters with larger proportions of young users had more significant use changes from $P_1$ to $P_2$. Therefore, we supposed that there was a certain degree of correlation between the user's usage changes during the pandemic and the user's age.

\subsection*{Discovering the main relevant factors of usage change patterns}

As we observed in the previous subsection, the changes in the daily usage of smartphones during the pandemic varied and were grouped into 9 clusters.
We also observed whether each cluster could be characterized by the distribution of the age of users and by the usage of apps in the pre-pandemic period.
If we characterized each cluster by this information, by gathering users who may belong to $C^{+, -}$ for example, we observed whether these people maintained stay-at-home orders from the increments of their smartphone usage in a period during which the government declared a state of emergency again.
Hence, in this subsection, we tried to characterize each cluster. 
We found the main relevant factors in each user cluster by establishing prediction models such as logistic regression, decision trees, and random forests that decide whether a user belongs to a cluster. That is, binary classification models were established.

We employed the distribution of age as an independent variable in the analysis, as shown in Fig \ref{Fig5}, and the usage of apps during $P_1$, as shown in Fig \ref{Fig6}. 
To represent the age range, we classified age into 6 groups every 10 years old as shown in Fig \ref{Fig5} and adopted 6 dummy variables for age groups.
For the use of apps during $P_1$, we used standardized data on the number of accesses of any target apps in each of 6 genres during each span of 4 time spans defined in the previous subsection.
Therefore, we used 6 dummies and 24 numerical independent variables.

\begin{figure}[!ht]
\includegraphics[scale=1]{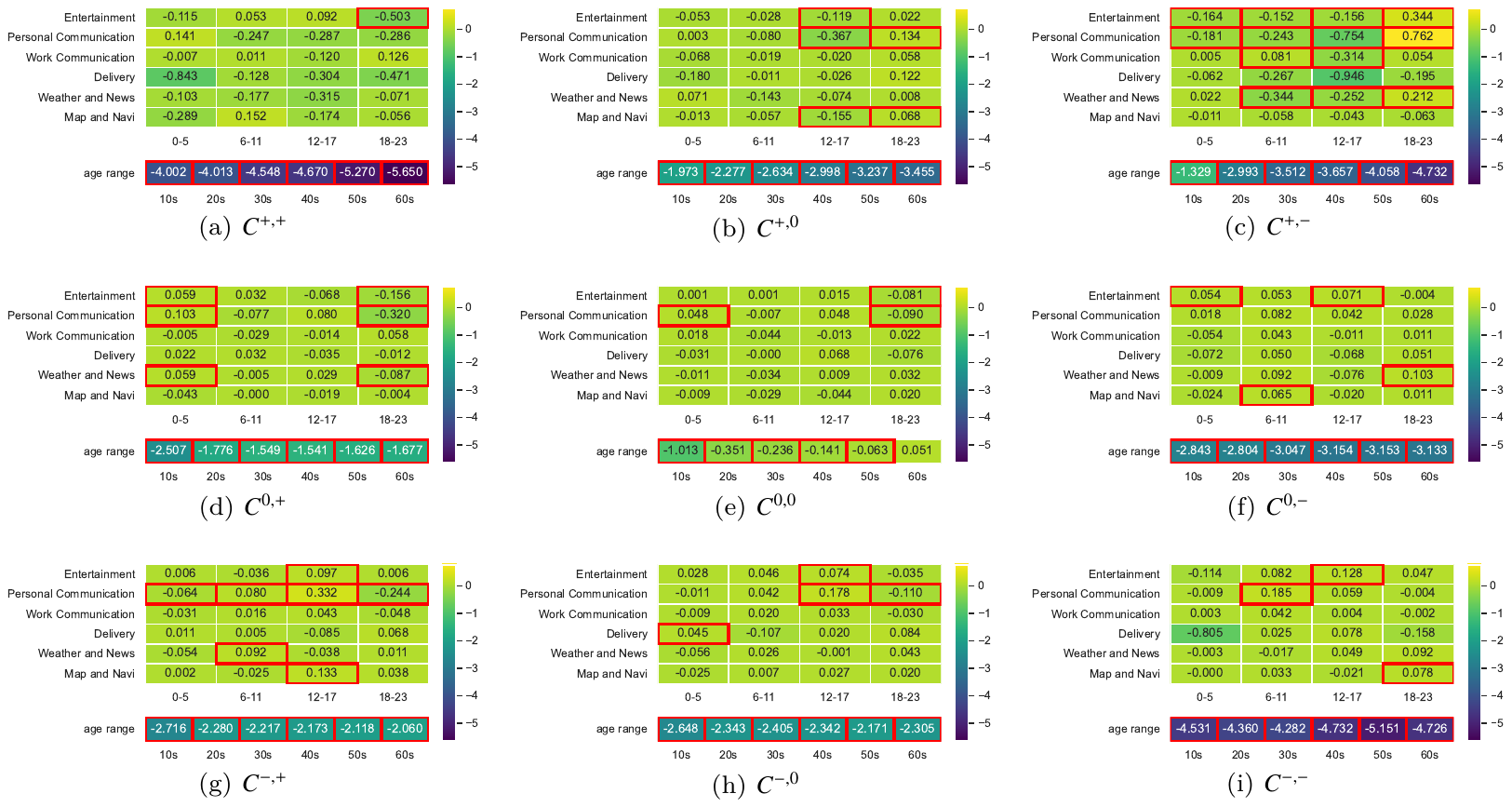}
\centering
\caption{{\bf Regression coefficient in the logistic regression model that characterize each cluster}}
\label{Fig7}
\end{figure}

The variable coefficients in the constructed logistic regression model, called the regression coefficients, can be interpreted as the change in the dependent variable when the corresponding independent variable changed while the other independent variables in the model remain constant.
Here, an independent variable with a large absolute value for its coefficient can be regarded as having a large effect on deciding whether the user belongs to the cluster or not.
In addition, a larger P-value indicates that the variable has a smaller effect on whether it belongs to the usage change cluster and usually the significance level (expressed as $\alpha$) is 0.05. 
Therefore, we show the regression coefficients of our model in Fig \ref{Fig7} and mark the value of variables with P-values below 0.05 by red borders, which means that the influence of variables on whether a user belongs to each usage change cluster is statistically significant.

\begin{figure}[!ht]
\includegraphics[scale=1]{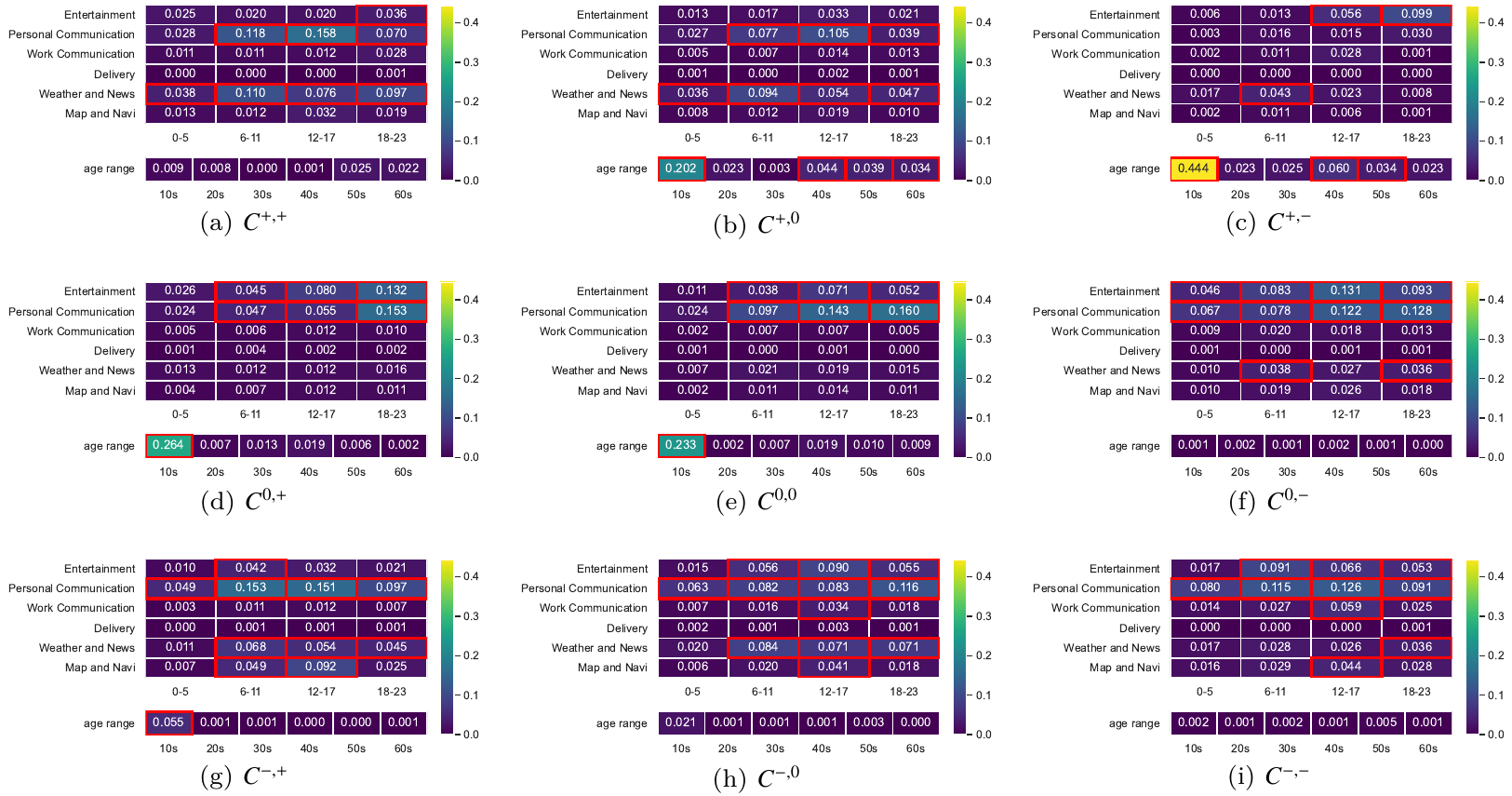}
\centering
\caption{{\bf Importance in the decision tree model for characterizing each cluster}}
\label{Fig8}
\end{figure}

\begin{figure}[!ht]
\includegraphics[scale=1]{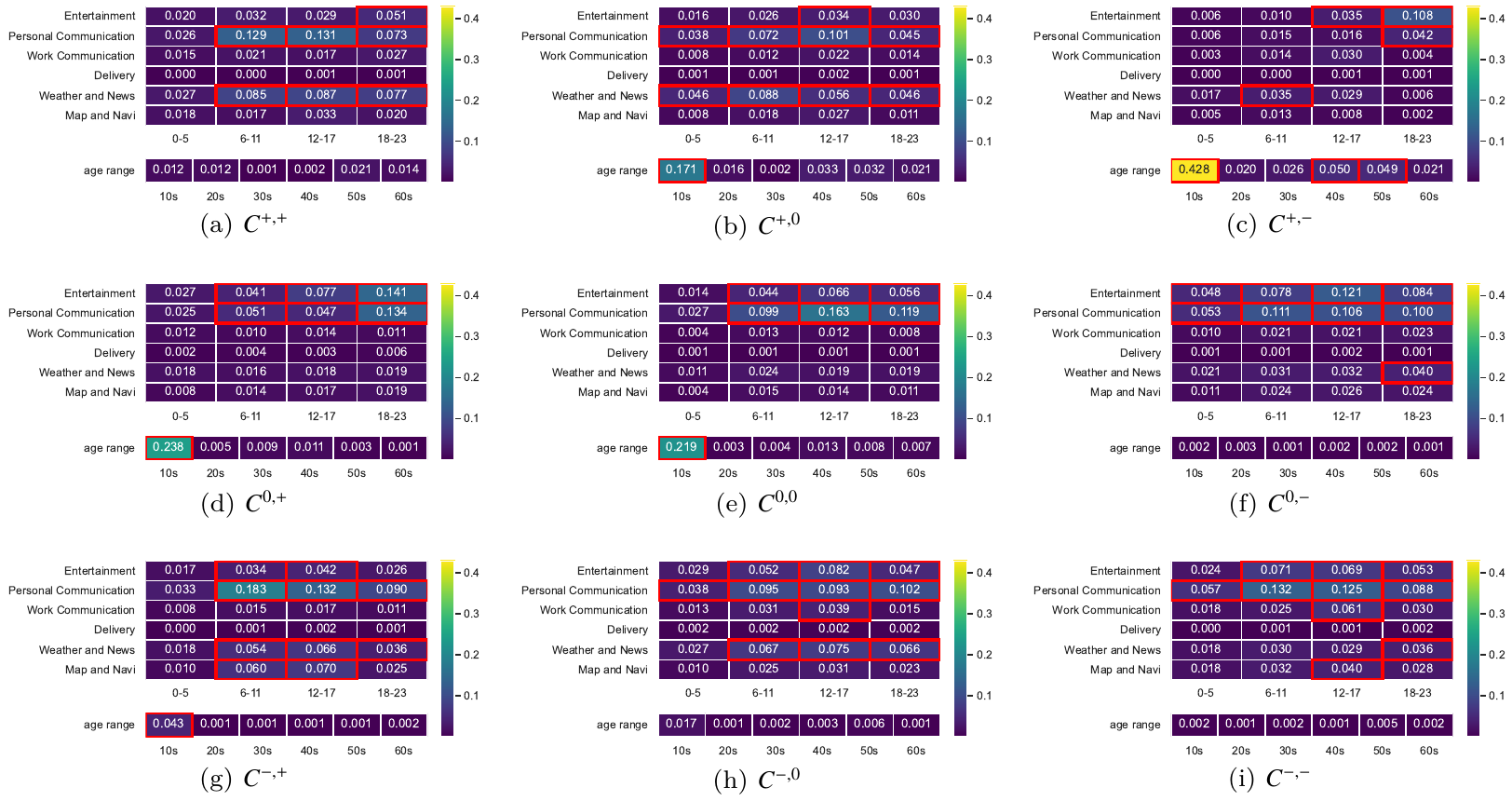}
\centering
\caption{{\bf Importance in the random forest model for characterizing each cluster}}
\label{Fig9}
\end{figure}

Constructing a decision tree model, we used the CART algorithm with a small amount of calculation, and the maximum depth of the tree was set to 5.
In the random forest model, we also used the CART algorithm and set the maximum depth of the tree to 5. The number of trees was set as 2700. 
The importance of each variable in the CART algorithm, which was measured by calculating the decrease in Gini impurity, is interpreted as the level at which the determination of each variable reduces the uncertainty in determining whether it belongs to each cluster. 
That is, it shows which variables are relevant to making a cluster.
To emphasize the variables with higher importance, we added red borders to variables whose importance exceeded the average value. 
In addition, in the algorithms we used for these models, the random number affected the selection of branch features. Therefore, we executed algorithms 50 times by selecting random seeds from 0 to 10000. In the results shown in Figs \ref{Fig8} and \ref{Fig9}, the average value of each variable's importance in 50 executions is shown. 
Figs \ref{Fig8} and \ref{Fig9} show that the rough trend in variable importance for every cluster in the decision tree model was similar to one in the random forest model.

As we observed in Fig \ref{Fig5}, 
the clusters with large increases in usage from $ P_1 $ to $ P_2 $ tended to also be younger age ranges, especially teens.
This fact can be confirmed from the regression coefficients of the logistic regression model and the variable importance of the decision tree and random forest models.
In the logistic regression model, the regression coefficients of age ranges for $ C ^ {+, +} $, $ C ^ {+, 0} $, and $ C ^ {+,-} $ were all negative, 
and these coefficients tended to decrease from teens to older adults. Users who were younger were more likely to be in these clusters than older users. 
The variable importance of the decision tree and random forest models also indicated that the percentage of users had a large influence on $ C ^ {+,0} $, $ C ^ {+,-} $, $C^{0,+}$, $C^{0,0}$, and $C^{-,+}$. 
For the clusters with a small change in usage such as $C^{0, +}$, $C^{0, 0}$, and $C^{-,+}$ combined with regression coefficients for teens, we found that compared to the users of other age ranges, young users were less likely to belong to these clusters. 
Therefore, being young was also an important factor in determining whether users belonged to these 2 clusters. This finding implied that the percentage of teen users had a strong effect on smartphone usage change patterns during the pandemic, whereas the percentage of other age ranges did not have a significant effect.

Combined with the regression coefficient, most of the usage change patterns were affected by 'Personal Communication' usage between 18 and 23 before the pandemic. 
In particular, according to the positive or negative regression coefficient, we found that the daily usage of those with more personal communication before the pandemic was more likely significantly increase during the self-restraint period.
In addition, the importance of 'Entertainment' usage between 18 and 23 before the pandemic was relatively high in clusters $C^{+,-}$, $C^{0,+}$, and $C^{0,0}$.
According to the logistic regression model, we found that when the use of 'Entertainment' between18 and 23 before the pandemic was higher, there was a higher probability that the user belonged to the cluster with a significant increase from $P_1$ to $P_2$, but it was less likely to belong to the clusters that remained unchanged from $P_1$ to $P_2$.

Furthermore, the importance of 'Personal Communication' use in the afternoon before the pandemic was relatively high in clusters $C^{+,0}$, $C^{-,+}$, and $C^{-,0}$. 
According to the regression coefficient, we deduced that the daily use of users who used more 'Personal Communication' between 12 and 17 before the pandemic had a tendency to decrease during the self-restraint period.
The importance of 'Weather and News' use in time spans 6--11 before the pandemic was relatively high in clusters $C^{+,-}$ and $C^{-,+}$. 
According to the regression coefficient, we inferred that users who used more 'Weather and News' in the morning hours before the pandemic had a tendency to reduce their daily use during the self-restraint period.

\section*{Conclusion}
Due to the quarantine policy during the pandemic, people's lifestyles and habits changed. This article analyzed the changes in people's behaviors from the perspective of the usage of smartphones.
First, we defined daily usage features based on the access log and showed daily usage change trends overall and in each genre of all users during the pandemic. 
By observing the overall situation in which people's behavior changed during the self-restraint period, we could understand and analyze the effects of restrictive measures and policies during the pandemic.  It may help to adopt better restrictive measures in the future. 
Next, we divided users into 9 types according to their smartphone usage changes during the three periods of the pandemic by clustering, which showed the diversity of people's usage changes during the pandemic.
Finally, according to the logistic regression, decision tree, and random forest models, we determined the influence of each age range and the usage of each genre and time span on the possibility of being a member of each cluster during the pandemic. This study validated the hypothesis that the age of users and daily smartphone usage during the period before the pandemic was related to the usage change patterns during the pandemic.
It could help to predict changes in smartphone actions during future pandemics or when other restrictive measures are implemented, which may become a new indicator to judge and manage the risks of measures or events according to the user's daily smartphone usage and demographic information.

The limitations of this study and future work are as follows.
First, although the COVID-19 pandemic has been ongoing, it is affected by data updates, and we were not able to analyze changes in usage after September 30, 2020. 
We will conduct follow-up research after we obtain the updated data.
Second, the prediction accuracy of the models we used to characterize the types of usage changes was not highly precise. 
Although the purpose was not to predict the change type, we hope to make a prediction model that can characterize the user's usage changes more accurately so that the correlation of the features we obtain will be more convincing.
In addition, this study did not have a more detailed perspective of the user's app usage pattern, e.g., the order of accessing the apps, to observe changes among users during the pandemic. 
In future work, we want to use patterns that often appear in daily use to observe and analyze the changes that occurred during the pandemic.

\section*{Acknowledgments}
We would like to thank Yasuaki Ohno of Fuller, Inc. for valuable discussion.



\end{document}